\documentclass[twocolumn,preprintnumbers,amsmath,prc,amssymb]{revtex4}

\usepackage{graphicx}
\usepackage{dcolumn}
\usepackage{bm}

\begin{document}

\title{Exploring an Origin of the QCD Critical Endpoint}

\author{K. A. Bugaev, V. K. Petrov  and G. M. Zinovjev}
\affiliation{Bogolyubov Institute for Theoretical Physics,
National Academy of Sciences of Ukraine, UA-03680, Kiev-143, UKRAINE
}

\date{\today}
\begin{abstract} 
We discuss a new way to develop the exactly solvable model of the QCD 
critical endpoint by matching the deconfinement phase transition line 
for the system of quark-gluon bags with the line of their vanishing 
surface tension coefficient. In contrast to all previous findings in 
such models the deconfined phase is defined not by an essential singularity 
of the isobaric partition function, but by its simple pole. As a result we 
find out that the first order deconfinement phase transition which is 
defined by a discontinuity of the first derivative of system pressure is 
generated by a discontinuity of the derivative of surface tension coefficient. 

\vspace*{0.5cm} 


\end{abstract}

\maketitle


\section{Introduction}

\vspace*{-0.25cm}

Recently intensive theoretical and experimental search for the (tri-)critical endpoint of 
strongly interacting matter at small enough chemical potential become very fascinating and 
promising branches of research activity in the context of relativistic heavy ion programs  
of many laboratories. In particular, the most powerful computers and very sophisticated 
algorithms are used for the lattice quantum chromodynamics (LQCD) simulations to locate 
this endpoint with maximal accuracy and to study its origin and properties \cite{fodorkatz,
karsch, deforcrandphilipsen}, but despite these efforts the present results are still far 
from being conclusive. The general arguments for a similarity with the critical point 
features in the other substances are not much convincing mainly because of a lack of 
rigorous critical point theory which exists, in a sense, for the spin systems only 
\cite{Wilson:eps}, whereas the origin and physics of the critical point for realistic 
gases and nuclear matter are described, at best, phenomenologically. For example, the 
Fisher droplet model (FDM) \cite{Fisher:67, Elliott:06} turns out rather efficient in 
studying the critical point of realistic gases. This model was applied to many different 
systems with the different extents of success including a nuclear multifragmentation  
\cite{Moretto}, a nucleation of real fluids \cite{Dillmann}, the compressibility factor 
of real fluids \cite{Kiang}, the clusters of the Ising model \cite{Ising:clust,Complement} 
and the percolation clusters \cite{Percolation}, but really its phase diagram does not 
include the fluid at all and, therefore, is not completely satisfactory and theoretically 
well defined. 

The statistical multifragmentation model (SMM) \cite{Bondorf:95} looks much more elaborated
in this aspect because defines the phase diagram of the nuclear liquid-vapor type phase 
transition (PT) in some controlled approximation \cite{simpleSMM:1, Bugaev:00} and 
predicts the critical (tri-critical) endpoint existence for the Fisher exponent 
$0 < \tau \le 1$ ($1 <  \tau \le  2 $) \cite{Bugaev:00} together with giving the 
possibility to calculate the corresponding critical exponents \cite{Reuter:01}. However, 
the predicted location of the SMM (tri-)critical endpoint at maximal density of the nuclear
liquid does not seem to be quite realistic. Actually, the relations between all these 
critical points are not well established yet. In principle, the Complement method 
\cite{Complement} provides us with the possibility to  describe accurately the size 
distribution of large clusters of 2- and 3-dimensional Ising model within the FDM in rather
wide temperature interval but the detailed numerical comparison of the Ising model and the 
FDM critical endpoints is hardly possible because of the large fluctuations even in 
relatively small systems. In meantime, taking the formal limit of the vanishing nucleon 
proper volume leads to the situation in which the SMM grand canonical partition function 
covers the FDM partition function, but the analytical properties are not the same and, as 
a result, the condensation  particle density of gaseous phase is finite not for the Fisher 
exponent $\tau \le  2$, as in the SMM, but for  $\tau > 2$ \cite{Reuter:01}. Besides, it 
leads to the various correlations between the $\tau$ exponent and other critical indices
in the FDM \cite{Fisher:67} and the corresponding relations in the SMM are different what 
signals the universality classes for these models are different as well \cite{Reuter:01}.

As to the model calculations of the QCD phase structure they are based on the universality 
arguments advanced in \cite{Rob:84} and concern mainly the temperature driven chiral 
symmetry restoration transition. In fact, it can not provide the reliable conclusion about
the transition order at $\mu$ = 0, its dependence on the number of flavors \cite{Misha:07}
and especially about the location of the point (tri-critical) on the PT line where the 
transition changes its order. It seems the lattice QCD (LQCD) simulations at vanishing $\mu$ 
give more definite evidence that this temperature driven phenomenon could really be a crossover 
\cite{aoki}. Then, clearly the $(\mu,T)$ phase diagram contains a critical point caused by 
the $\mu = 0$ crossover turning into the first order PT. However, this wisdom is rather
questionable as well \cite{adrianodig}.

Furthermore, the recent LQCD simulations \cite{fodorkatz,karsch,LQCD:rev} teach us that  
even at high temperatures up to a few $T_c$ ($T_c$ is the cross-over temperature), a QGP does 
not consist of the weakly interacting quarks and gluons and its pressure and energy density 
are well below of the corresponding quantities of non-interacting quarks and gluons. Although 
such a strongly coupled QGP (sQGP) \cite{Shuryak:sQGP} has put a new framework for the QCD 
phenomenology, the feasibility of understanding such a behavior within the AdS/CFT 
duality\cite{AdS} or statistical approaches is far from being simple and transparent. 

Here we investigate the possibility to resolve the problem by formulating an approach based
on the model of quark-gluon bags with surface tension (QGBSTM) \cite{QGBSTM,QGBSTM:2,FWM:08}.
The paper is organized as follows. Sect. II contains the formulation of model basic elements 
(hereafter the model is named  QGBSTM2 in order to distinguish it from the model with the 
tri-critical endpoint). In Sect.III we analyze all possible singularities of the QGBSTM2 
isobaric partition for non-vanishing baryonic densities and discuss the necessary conditions 
for the critical point existence. The conclusion  are summarized in 
Sect.IV. 


\section{Model of quark-gluon bags with surface tension}


The most convenient way to study the phase structure of the QGBSTM is to use the isobaric 
partition \cite{QGBSTM, Bugaev:05c} analyzing its rightmost singularities. Hence, we assume 
that after the Laplace transform the QGBSTM2 grand canonical partition $Z(V,T,\mu)$ generates 
the following isobaric one:
\vspace*{-0.05cm}
\begin{eqnarray}\label{Zs}
\hspace*{0.05cm}\hat{Z}(s,T,\mu) \equiv \hspace*{-.05cm} \int\limits_0^{\infty}\hspace*{-.05cm}dV\, e^{\textstyle -sV}\,Z(V,T,\mu) =\frac{1}{ [ s - F(s, T,\mu) ] } \,,
\end{eqnarray}

\vspace*{-0.2cm}
\noindent
where the function $F(s,T,\mu)$ includes \cite{QGBSTM} the discrete $F_H$ and continuous 
$F_Q$ volume spectra of the bags  
%
%
\begin{eqnarray}
\hspace*{-0.4cm}  F(s,T,\mu) &\hspace*{-0.1cm}\equiv& \hspace*{-0.1cm}F_H(s,T,\mu)+F_Q(s,T,\mu) =  \nonumber  \\
&\hspace*{-0.85cm}= &\hspace*{-0.45cm}\sum_{j=1}^n g_j e^{\textstyle (\frac{\mu}{T}b_j -v_js)} \phi(T,m_j) +  \\
\label{FsH}
\hspace*{-0.6cm}&\hspace*{-0.75cm}+&  \hspace*{-0.45cm} {\textstyle u(T)} \hspace*{-0.1cm} \int\limits_{V_0}^{\infty} \hspace*{-0.1cm}\frac{dv}{v^{\tau}}\hspace*{0.1cm} 
e^{\textstyle [ \left(s_Q(T,\mu)-s \right)v - \Sigma(T,\mu) 
v^{\varkappa}]  }\, .
\label{FsQ}
\end{eqnarray}
\noindent
$u(T)$ and $s_Q(T,\mu)$ are continuous and, at least, double differentiable functions of their 
arguments (see \cite{QGBSTM,FWM:08} for details). 
The density of bags having mass $m_k$, eigen volume $v_k$, baryon charge $b_k$ and degeneracy $g_k$
is given by  $\phi_k(T) \equiv g_k ~ \phi(T,m_k) $  with 
\vspace*{-0.cm}
\begin{eqnarray} 
\phi_k(T)   & \equiv  \frac{g_k}{2\pi^2} \int\limits_0^{\infty}\hspace*{-0.0cm}p^2dp~
\exp{\textstyle \left[- \frac{(p^2~+~m_k^2)^{1/2}}{T} \right] } =
\nonumber \\
& =  g_k \frac{m_k^2T}{2\pi^2}~{ K}_2 {\textstyle \left( \frac{m_k}{T} \right) }\, .
\end{eqnarray}
The continuous part of the volume spectrum (\ref{FsQ}) is a generalization of exponential mass 
spectrum introduced by Hagedorn \cite{Hagedorn:65} and it can be steadily derived in both the MIT 
bag model \cite{Kapusta:81} and finite width QGP bag model \cite{FWM:08}. The term $e^{-s v}$ 
describes the hard-core repulsion of the Van der Waals type. $\Sigma(T,\mu) $ denotes the ratio  
between the $T$ and $\mu$ dependent surface tension coefficient and $T$ (the reduced surface 
tension coefficient hereafter) which has the form 
\begin{eqnarray}\label{Sigma}
\Sigma(T, \mu) =  
\left\{ \begin{array}{rr}
\Sigma^- > 0  \,,  &\hspace*{0.1cm}  
T \rightarrow T_{\Sigma} (\mu)  - 0 \,,\\
 0 \,, &\hspace*{0.1cm}  T =  T_{\Sigma} (\mu) \,, \\
\Sigma^+ < 0 \,,  &\hspace*{0.1cm}
T \rightarrow T_{\Sigma} (\mu)  + 0  \,.
\end{array} \right. 
\end{eqnarray}
At making choice in favour of such a simple surface energy parameterization we 
follow the original Fisher idea \cite{Fisher:67} which allows one to account for 
the surface energy by considering a mean bag of volume $v$ and surface extent $v^{\varkappa}$.  
As it has been discussed in \cite{QGBSTM,QGBSTM:2} the power $\varkappa < 1$ inherent in 
bag effective surface is a constant which, in principle, is different from the typical FDM 
and SMM value $\frac{2}{3}$.

Let us stress here that we do not require the precise disappearance of $\Sigma(T,\mu)$ above the 
critical endpoint as it is usual in FDM and SMM. It was shown in \cite{QGBSTM} and is argued here,  
this point is found crucial in formulating the statistical model with deconfining cross-over
(in contrast with previous efforts \cite{Goren:05,Nonaka:05}). We would like also to note the
negative value of the reduced surface tension coefficient $\Sigma(T,\mu)$ above the 
$T_\Sigma (\mu)$-line in the $(\mu, T)$-plane should not be surprising. It is the well-known fact  
that in the grand canonical ensemble the surface tension coefficient includes the energy and 
entropy contributions which have the opposite signs \cite{Fisher:67,Elliott:06,Bugaev:04b}. 
Therefore, $\Sigma(T,\mu) < 0 $ does not mean that the surface energy changes the sign, but it
rather signals that the surface entropy contribution simply exceeds the surface energy part
and results in the negative values of surface free energy.  In other words, 
the number of non-spherical bags of fixed volumes becomes so big that the Boltzmann exponent
which accounts for the energy "costs" of these bags does not provide their suppression anymore.  
Such a situation is standard for the statistical ensembles with the fluctuating extensive 
characteristics (the surface of fixed volume bag fluctuates around the mean value)  \cite{FWM:08b}. 

\begin{figure}[ht]
\includegraphics[width=6.3cm,height=6.cm]{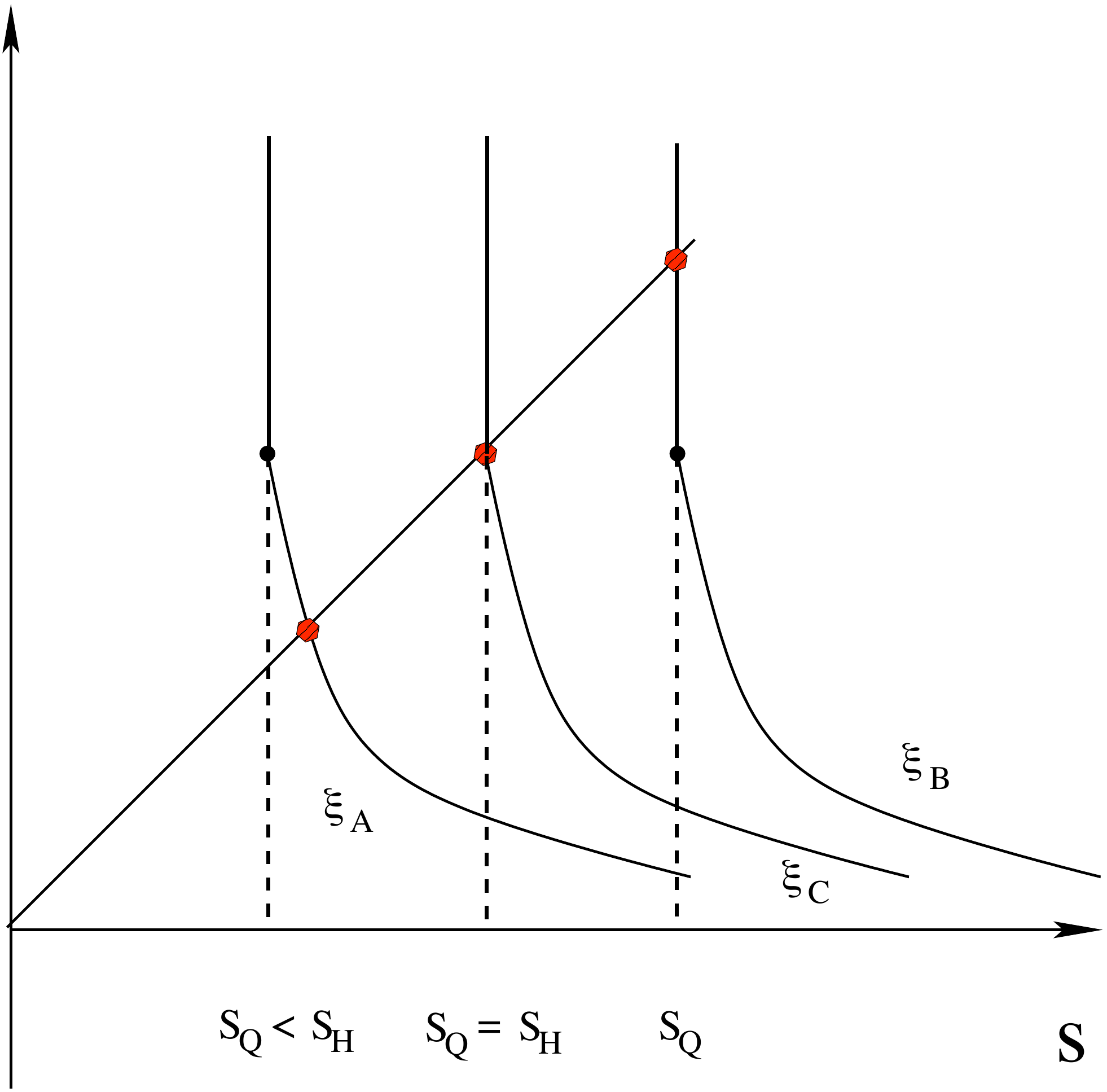}
\vspace*{-0.3cm}
\caption{[Color online]
Singularities of the isobaric partition (\ref{Zs}) and the corresponding 
graphical solution of Eq. (\ref{EqVI}) which describes a PT in the models similar to QGBSTM.
The solution of Eq. (\ref{EqVI}) is shown by a filled hexagon.
$F(s, \xi)$ is shown by a solid curve for a few values of the parameter sets $\xi$.  $F(s, \xi)$ diverges 
for $s < s_Q( \xi)$ (shown by dashed lines), but is finite at $s = s_Q( \xi)$ (shown by black circle).  
At low values of  the parameters  $\xi = \xi_A$, which can be either 
$\xi \equiv  \{T, \mu = const \} $ or $\xi \equiv \{ T = const, \mu\}$, 
the simple pole $s_H$ is the rightmost singularity and it corresponds to hadronic phase. 
For  $\xi = \xi_B \gg \xi_A$ the  rightmost singularity is an essential singularity $s = s_Q( \xi_B)$, 
which describes QGP. 
At intermediate  value $\xi = \xi_C$ both singularities coincide $s_H( \xi_C) = s_Q( \xi_C)$ and 
this condition is a Gibbs criterion (\ref{EqVII}). 
At transition from the low energy density phase to the high density one the rightmost singularity changes 
from the simple pole to the essential singularity.
}
  \label{fig1}
\end{figure}

By construction the isobaric partition (\ref{Zs}) develops two types of singularities:
the simple pole $s^* = s_H(T,\mu)$ which is defined by the equation
\begin{equation}\label{EqVI}
s^*~=~ F(s^*,T,\mu) \,,
\end{equation}
and in addition there appears an essential singularity $s^* = s_Q(T,\mu)$ which is defined by 
the point $s = s_Q(T,\mu) -0$ where the continuous part of spectrum $F_Q(s,T,\mu)$ (\ref{FsQ})
becomes divergent. This singularity is also defined by Eq.(\ref {EqVI}). 
Usually the statistical models similar to QGBSTM \cite{CGreiner:06,Bugaev:05c, QGBSTM}
have the following structure of singularities. The pressure of low energy density phase (confined) 
$p_H (T,\mu)$ is described by the simple pole $s = s_H(T,\mu) = \frac{p_H (T,\mu)}{T}$ which is 
the rightmost singularity of the isobaric partition (\ref{Zs}), whereas the pressure of high energy 
density phase (deconfined) $p_Q (T,\mu)$ defines the system's pressure, if the  essential 
singularity  $s = s_Q(T,\mu) = \frac{p_Q (T,\mu)}{T}$ of this partition becomes the rightmost one 
(see Fig.\ref{fig1}). Such an interplay of rightmost isobaric partition singularity and the 
pressure of the grand canonical ensemble is the typical feature of the Laplace transform technique
\cite{CGreiner:06,Bugaev:05c}. 

The deconfinement PT occurs at the equilibrium line $T_c(\mu)$ where both singularities match 
each other
\begin{equation}\label{EqVII}
s_H(T, \mu) ~=~ s_Q(T, \mu)  \quad \Rightarrow  \quad T = T_c(\mu)  \,.
\end{equation}
In this equation one can easily recognize the Gibbs criterion for phase equilibrium.  
Such a behavior of the rightmost singularities is shown in Fig.1. 

It was demonstrated in \cite{QGBSTM} the deconfinement PT takes place if the phase equilibrium  
temperature (\ref{EqVII}) is lower than the  temperature of the null surface tension line (\ref{Sigma}) 
for the same value of baryonic chemical potential, i.e. $T_c(\mu) < T_{\Sigma}(\mu)$, whereas at low 
values of $\mu$ the PT is degenerated into a cross-over because the line $T = T_{\Sigma} (\mu)$ leaves 
the QGP phase to appear in the hadronic phase. The intersection point $(\mu_{end};T_c(\mu_{end}))$ of 
these two lines $T_c(\mu) = T_{\Sigma}(\mu)$ is the tricritical endpoint \cite{QGBSTM} since 
for $\mu \ge \mu_{end}$ and $T > T_c(\mu_{end})$ at the null surface tension line $T = T_{\Sigma} (\mu)$  
there exists the surface induced PT \cite{QGBSTM}. 

The important element of our deliberation here is a way found out to get rid of the surface induced PT  
and to `hide' it inside the deconfining one. In order to demonstrate the result we assume the surface 
tension coefficient changes its sign exactly at the deconfinement PT line, i.e. 
for $\max\{\mu (T_c)\}  \ge \mu \ge \mu_{end}$ and $T \le T_c(\mu_{end})$ one has  
$T_c(\mu) = T_{\Sigma}(\mu)$ while keeping the cross-over transition for $\mu < \mu_{end}$ similar to
\cite{QGBSTM}. The possibility to match these two PT lines was clear long ago, but a nontriviality is
seen in the fact that an existence of both the critical endpoint at $(\mu_{end}; T_c(\mu_{end}))$ 
and the 1$^{st}$ order deconfinement PT  at  $ T_c(\mu)  =  T_{\Sigma} (\mu)$ is generated by an entire  
change of the rightmost singularity pattern. 

%
%

\vspace*{-0.0cm}

\vspace*{-0.cm}

\section{Conditions for the critical endpoint existence}

\vspace*{-0.0cm}

Under adopted assumption the rightmost singularity in the QGBSTM2 is always the simple pole since 
in the right hand side vicinity of $s \rightarrow s_Q(T, \mu) + 0$ the value of 
$F_Q(s,T,\mu) \rightarrow \infty $ for $\Sigma = \Sigma^+ < 0$. Then the motion of singularities 
corresponds to Fig. 2 in this situation. The question, however, appears whether such a behavior 
corresponds to PT. To clarify the point it is convenient to introduce the variable 
$\Delta^\pm \equiv \Delta (T_\Sigma \pm 0, \mu) = s^\pm  - s_Q(T_\Sigma \pm 0, \mu)$ and to compare 
the $T$ derivative of the right most singularity $s^- \equiv s^*(T_\Sigma - 0, \mu)$  below and   
$s^+ \equiv s^*(T_\Sigma + 0, \mu)$ above the PT line $ T_c(\mu) = T_{\Sigma} (\mu) $ for the same
magnitudes of $\mu$. Due to the relation between the system pressure $p(T,\mu)$ and the rightmost 
singularity $s^*(T, \mu) = \frac{p(T,\mu)}{T}$, the difference of $T$ derivatives, 
$\frac{\partial (\Delta^+ - \Delta^-) }{\partial~~ T}$, if revealed on both sides of the PT line  
is defined by the difference of the corresponding entropy densities. Therefore, according to the 
standard classification of the PT order an appearance of nonzero values of 
$\frac{\partial (\Delta^+ - \Delta^-) }{\partial~~ T} \neq 0$  signals  about the 1$^{st}$ order PT.

%
%
\begin{figure}[ht]
\includegraphics[width=6.3cm,height=6.cm]{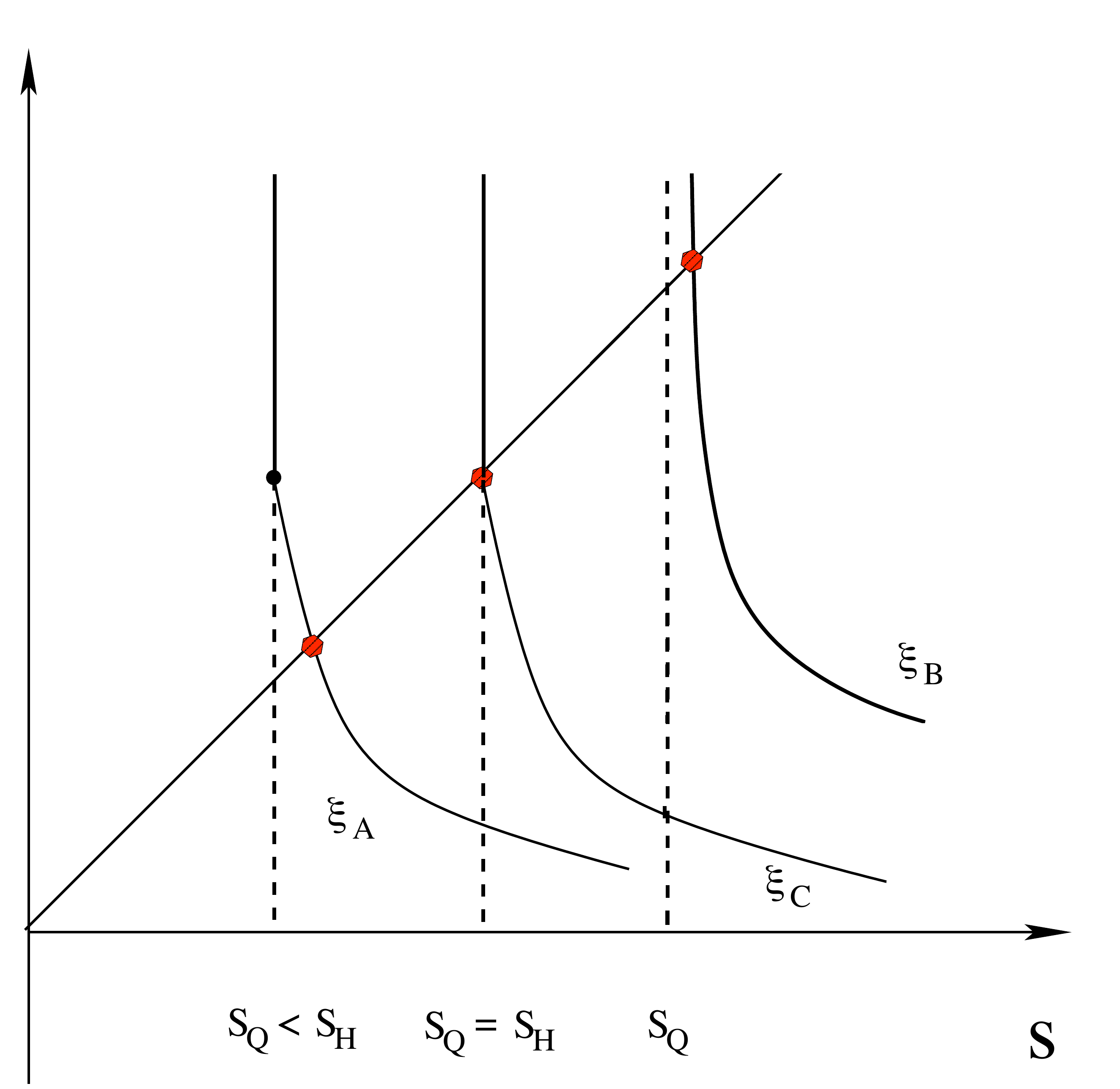}
\vspace*{-0.3cm}
\caption{[Color online]
Singularities of the isobaric partition (\ref{Zs}) and the corresponding 
graphical solution of Eq. (\ref{EqVI}) which describes a PT in the  QGBSTM2.
The legend corresponds to Fig. \ref{fig1}.
In this case, however, 
the rightmost singularity for each phase is the simple pole, whereas at the PT the essential 
singularity matches the simple pole due to the vanishing surface tension coefficient.  
}
  \label{fig2}
\end{figure}

\vspace*{-0.0cm}

Now using the auxiliary functions 
\begin{eqnarray}\label{EqVIII}
%
{\cal K}_{a} \left(x  \right) & \equiv&  \hspace*{-0.15cm} 
 \int\limits_{V_0 \Delta}^\infty  \hspace*{-0.15cm}
dz ~\frac{\exp\left[- z  + x  z^{\varkappa} 
\right] }{z^{a}}   \,, \\
g_\tau(\Delta^\pm, \Sigma^\pm) & \equiv & \frac{ \exp\left[ -\Delta^\pm V_0 - 
\Sigma^\pm V_0^{\varkappa}\right] }{(\tau-1)\, V_0^{\tau-1}} \,, 
\end{eqnarray}\label{EqIX}
\noindent
it is possible to rewrite the continuous part of volume spectrum (\ref{FsQ}) as
$F_Q(s^\pm,T,\mu) = {\textstyle u(T)} I_\tau (\Delta^\pm, \Sigma^\pm)$
integrating by parts the following integral  
\begin{align}\label{EqX}
 &I_\tau (\Delta^\pm, \Sigma^\pm) \equiv     \hspace*{-0.05cm} 
 \int\limits_{V_0}^\infty  \hspace*{-0.05cm} d v~  \frac{\exp\left[- \Delta^\pm v  - \Sigma^\pm  v^{\varkappa} 
\right] }{v^{\tau}}   =   \nonumber \\
&   \left[ g_\tau(\Delta^\pm, \Sigma^\pm) - 
\frac{\Delta^\pm}{\tau-1} g_{\tau-1}(\Delta^\pm, \Sigma^\pm)  \right.      -   \nonumber \\
 &   \frac{ \varkappa \Sigma^\pm }{\tau-1}   \hspace*{-0.cm}g_{\tau-\varkappa }(\Delta^\pm, \Sigma^\pm) +  \left. 
\frac{(\Delta^\pm)^{\tau-1}}{\tau-1} \Phi \left(  -\frac{\Sigma^\pm}{ (\Delta^\pm)^\varkappa  } \right)   \right] \,, \\
& \Phi ( x )   \equiv    {\cal K}_{\tau-2} (x ) - 
\frac{\varkappa(2\tau -3 -\varkappa)\, x }{(\tau-2)(\tau-1-\varkappa)} {\cal K}_{\tau-1-\varkappa} (x ) +
 \nonumber \\ 
&
\frac{\varkappa^2 \, x^2}{\tau-1-\varkappa} {\cal K}_{\tau -2 \varkappa} (x ) \,.
\label{EqXI}
 \end{align}
Drawing Eqs. (\ref{EqX}) and (\ref{EqXI}) one can show the necessary condition of deconfinement PT  
existence at $\Sigma^\pm \rightarrow 0 $ becomes  
$s_Q(T_\Sigma, \mu) = F_H (s_Q(T_\Sigma, \mu), T_\Sigma, \mu ) + u(T_\Sigma)  g_\tau (0,0)$
and it provides $\Delta^\pm  \rightarrow  +0 $, indeed. For $\tau < 1 + 2 \varkappa$ such a statement follows 
directly from the present form of (\ref{EqXI}), whereas for larger values of $\tau$ exponent one needs to 
integrate ${\cal K}_{a} (x)$-functions in (\ref{EqXI}) while they converge at the lower integration limit 
for $\Delta^\pm\rightarrow  +0$.
  
With treating Eqs. (\ref{EqVIII})-(\ref{EqXI}) one can easily find 
\begin{align}\label{EqXII}
\frac{\partial \Delta^\pm }{\partial ~T} = & \frac{  \frac{\partial F_H}{\partial T}  + 
\frac{\partial s_Q}{\partial T}\left[\frac{\partial F_H}{\partial s} - 1 \right] + \frac{\partial u}{\partial T} I_\tau (\Delta^\pm, \Sigma^\pm)  }{ 1 + u I_{\tau-1} (\Delta^\pm, \Sigma^\pm) -  \frac{\partial F_H}{\partial s}  }  - \nonumber \\
& \frac{u I_{\tau-\varkappa}(\Delta^\pm, \Sigma^\pm) \frac{\partial \Sigma^\pm}{\partial T} }{1 + u I_{\tau-1}(\Delta^\pm,\Sigma^\pm) -  \frac{\partial F_H}{\partial s}  }  \,,
 \end{align}
which in the limit $\Delta^\pm, \Sigma^\pm \rightarrow 0 $ gives 
\begin{align}\label{EqXIII}
\frac{\partial \Delta^+ }{\partial ~T} - \frac{\partial \Delta^- }{\partial ~T}  \rightarrow & -  
\frac{ u I_{\tau-\varkappa} (0,0) \left[\frac{\partial \Sigma^+}{\partial T} - \frac{\partial \Sigma^- }{\partial T}\right]  }{ 1 + u I_{\tau-1} (0,0) -  \frac{\partial F_H}{\partial s}  }   \,. 
 \end{align}
This is a remarkable result because it clearly shows in the present model the 1$^{st}$ order deconfinement 
PT does exist, if the $T$ derivative of reduced surface tension coefficient has a discontinuity 
at the phase equilibrium line only!
Thus, a discontinuity of the first derivative of a system pressure, which is a three-dimensional quantity, 
is generated by a discontinuity of the derivative of surface tension coefficient, which is a two-dimensional 
characteristics. In the other words, within the QGBSTM2 the deconfinement 1$^{st}$ order PT is just a surface 
induced one. The necessary condition for its existence is the finiteness of integrals 
$I_{\tau-\varkappa}(0,0)$ and $I_{\tau-1}(0,0)$ in (\ref{EqXIII}), i.e. $\tau > 2$.

Moreover, to realize a PT from hadronic matter to QGP it is necessary to have at the PT line   
$\frac{\partial \Delta^+ }{\partial ~T} - \frac{\partial \Delta^- }{\partial ~T} =   
\frac{1}{T}\frac{\partial}{\partial T} \left[ p_Q(T,\mu) - p_H(T,\mu)\right] > 0$ and, hence, at this line 
\begin{align}\label{EqXIV}
 \frac{\partial \Sigma^+}{\partial T}  -  \frac{\partial \Sigma^- }{\partial T}  < 0 \,.
 \end{align}

Now it is clear that at the critical endpoint $(\mu_{end};T_c(\mu_{end}))$ the entropy density gap vanishes 
due to the disappearing difference $\frac{\partial\Sigma^+}{\partial T} - \frac{\partial\Sigma^- }{\partial T} = 0$. 

With the general parameterization of reduced surface tension coefficient which is consistent with (\ref{Sigma})
\begin{eqnarray}\label{EqXV}
\Sigma(T, \mu) =  \frac{1}{T} \cdot
\left\{ \begin{array}{rr}
\sigma^- \left[ \frac{ T_{\Sigma}  (\mu)  - T }{T_{\Sigma}  (\mu) } \right]^{\zeta^-} \,,  &\hspace*{0.1cm}  
T \rightarrow T_{\Sigma} (\mu)  - 0 \,,\\
 & \\
- \sigma^+ \left[ \frac{ T - T_{\Sigma} (\mu)   }{T_{\Sigma}  (\mu) } \right]^{\zeta^+} \,,  &\hspace*{0.1cm}
T \rightarrow T_{\Sigma} (\mu)  + 0  \,,
\end{array} \right. 
\end{eqnarray}
we are able to conclude about the powers $\zeta^\pm$ and the values of coefficients $\sigma^\pm \ge  0$.  
It is obvious from (\ref{EqXII}) that $\zeta^\pm \ge 1$, otherwise the corresponding entropy density is 
divergent at the PT line. If, for instance,  $\zeta^+ = 1$, as predicted by the Hills and Dales 
model \cite{Bugaev:04b}, then $\zeta^- =1 $, and  according to (\ref{EqXIV}) one has 
$\sigma^+ > \sigma^-$. If, however,  $\zeta^- > 1$, then  from (\ref{EqXIV}) it follows that 
$ \sigma^+ \, \zeta^+ (T - T_\Sigma (\mu))^{\zeta^+ - 1} > 0$ for $T \rightarrow T_\Sigma (\mu) + 0$.
The latter is consistent with the equality $\zeta^+ = 1$. 

It can be shown that in accordance with (\ref{EqVII}) the inequalities  
\begin{align}\label{EqXVI}
\hspace*{-0.35cm}{\textstyle \frac{\partial F_H}{\partial T}  + 
 \frac{\partial s_Q}{\partial T} \left[  \frac{\partial F_H}{\partial s}  - 1 \right]  + \frac{\partial u}{\partial T} 
 g_\tau (0, 0) \gtrless u g_{\tau-\varkappa}(0,0)\frac{\partial \Sigma^\pm}{\partial T}
}
 \end{align}
are the sufficient conditions of the 1$^{st}$ order PT existence that provide (\ref{EqXIV}) and guarantee 
the uniqueness of solutions $\Delta^\pm  \rightarrow  +0 $ on both sides of the PT line. 

The critical endpoint $(\mu_{end}; T_c(\mu_{end}))$ exists, if in its vicinity the difference of 
coefficients $\sigma^\pm$ vanishes as 
\begin{align}\label{EqXVII}
\sigma^+ - \sigma^- \sim d^{\zeta_{end}},~~ d \equiv T - T_c(\mu_{end}) - \frac{\partial T_\Sigma}{\partial \mu}\biggl|_{\mu_{end}} \hspace*{-0.4cm} (\mu - \mu_{end})  
 \end{align}
with $\zeta_{end} \ge 1$.
By construction in the $\mu-T$ plane $d$ as defined by (\ref{EqXVII}) vanishes at the tangent line to 
the PT curve at $(\mu_{end}; T_c(\mu_{end}))$. As one can easily  see from either $T$ or $\mu$ derivative 
of (\ref{EqXII}) any second derivative of the difference $\Delta^+ - \Delta^- = 0$ at the  critical endpoint 
$(\mu_{end}; T_c(\mu_{end}))$, if $\zeta^+ = \zeta^- = \zeta_{end} = 1$ only, which provides the 
2$^{nd}$ order PT available at this point. The higher order PT at the critical endpoint may exist for 
$ \zeta_{end} = 2$


\section{Conclusion}


Here we presented new exactly solvable model, QGBSTM2, (or even the class of models) which develops 
the critical endpoint at $(\mu_{end};T_c(\mu_{end}))$. This model naturally explains the transformation 
of the 1$^{st}$ order deconfining PT into a weaker PT at the endpoint and into a cross-over at low 
baryonic densities as driven by negative surface tension coefficient of the QGP bags at high energy 
densities. It sheds new light on the QGP equation of state suggested in Ref.\cite{QGBSTM} where it has 
been shown that the deconfined QGP phase presents itself just a single infinite bag whereas 
the cross-over phase consists of the QGP bags of all possible volumes and only at very high pressure
values this phase is presented by one large (infinite) bag. The important consequence of such a property 
is that the deconfined QGP phase should be separated from the cross-over QGP by another PT which is 
induced by the change of surface tension coefficient sign of large bags. Furthermore, QGBSTM teaches us 
that for the Fisher exponent the 1$^{st}$ order deconfinement PT exists for $1 < \tau \le 2$ only, 
whereas at the endpoint there exists the 2$^{nd}$ order PT for $\frac{3}{2} < \tau \le 2$ and this point 
is the tri-critical one. 

On the other hand the important message of QGBSTM2 is that a solvable model of the QCD critical endpoint 
can be formulated for $\tau >2$. Technically it is achieved by matching the deconfinement PT line with 
the line of vanishing surface tension coefficient $T_\Sigma(\mu)$ for $\mu \ge \mu_{end}$ and 
$T \le T_c(\mu_{end})$. This step leads to new strong assertion that the 1$^{st}$ order PT in QGBSTM2 is
not accompanied by change of the leading singularity type as was argued earlier in Refs.
\cite{CGreiner:06,Bugaev:05c}. Thus, the high density QGP phase is defined by not an essential singularity
of the isobaric partition (\ref{Zs}) but its simple pole. Similar to QGBSTM the high density phase of this 
model is defined by the QGP crossover whereas the deconfined matter (an interior of single infinite bag)
may exist at the mixed phase inherent in the deconfinement PT only.
Besides we find also that the 1$^{st}$ order deconfining PT, i.e. a discontinuity of the first derivative 
of a system pressure, which is a three-dimensional quantity, is generated by the discontinuity of surface 
tension coefficient derivative, which is a two-dimensional quantity. Thus, we explicitly show that within 
the  present model the deconfinement 1$^{st}$ order PT is the surface induced one. \\ 
\indent
Another distinctive  feature of these results is that for the first time we see the critical endpoint 
in the model with the constituents of nonzero proper volume exists not for $\tau \le 1$ as in the 
SMM \cite{simpleSMM:1, Bugaev:00} and not for $1< \tau \le 2$ as the tricritical endpoints in the SMM 
and in the QGBSTM \cite{QGBSTM}, but for $\tau >2$, i.e. as in FDM \cite{Fisher:67}. Perhaps, this feature  
may be helpful to distinguish experimentally the QCD critical endpoint from the tri-critical one.



\end{document}